\documentclass[aps,twocolumn,pra,superscriptaddress,amsmath,showpacs,tightenlines]{revtex4-1}
\usepackage{amssymb}
\usepackage{amsmath}
\usepackage{graphicx}
\usepackage{subfigure}
\usepackage{natbib}
\usepackage{epsfig}
\usepackage{amsfonts}
\usepackage{mathrsfs}
\usepackage{xcolor}
\usepackage[toc,page,title,titletoc,header]{appendix}
\begin{document}

\title{Retardation effect and dark state in a waveguide QED setup with rectangle cross section}
\author{Yang Xue}
\affiliation{National Demonstration Center for Experimental Physics Education,
Northeast Normal University, Changchun 130024, China}
\affiliation{Center for Quantum Sciences and School of Physics, Northeast Normal University, Changchun 130024, China}
\author{Zhihai Wang}
\email{wangzh761@nenu.edu.cn}
\affiliation{Center for Quantum Sciences and School of Physics, Northeast Normal University, Changchun 130024, China}

\begin{abstract}
In this paper, we investigate the dynamics of a two-atom system which couples to a quasi-one dimensional waveguide with rectangle cross section. The waveguide supports different TM and TE modes and the former ones play as environment by on-demand choosing the dipole moment of the atoms. Such environment induces the interaction and collective dissipation of the atoms. When both of the two atoms are located in the middle of the waveguide, we observe a retardation effect, which is broken by moving one of the atom to be off-centered. To preserve the complete dissipation of the system via dark state mechanism, we propose a scheme where the connection of the atoms are perpendicular to the axis of the waveguide. We hope our study will be useful in quantum information processing based on state-to-art waveguide structure.
\end{abstract}


\maketitle
\section{introduction}

The waveguide QED, which studies the light-matter interaction in a confined structure, has attracted  lots of attention due to its interesting theoretical and experimental applications~\cite{DR2017,XG2017}.
In the waveguide QED setup, how to control the photon by the (artificial) atom  and vise verse is a central task to construct the quantum network. On the one hand, the propagation of the flying photon can be controlled by the frequency of the atom, which is widely used to realize coherent quantum device such as photon transistor~\cite{JT2005,DE2007,LZ2008}, router~\cite{AA2010,IC2011,LZ2013,Wang2014,IS2014,CH2018,YL2022}, etc. On the other hand, the waveguide can serve as a data bus, to induce the interaction between different atoms~\cite{KS2012,AG2013,CG2013,HP2015,GC2016,AG2017,FG2018,HZ2019,EK2021}, which is utilized to realize remote quantum entanglement.

Due to the possible slow velocity of light in the waveguide, the time needed for the photon propagating from one atom to the other can be comparable to the lifetime of the atom. Therefore, the retardation effect, which will induce some non-Markovian dynamics, is becoming a hot topic recently. Such retardation effect will occur in multiple atoms system~\cite{FT1970,PW1974,Qu2012,HP2016,KS2020} or only one atom in front of a mirror~\cite{JE2001,UD2002,PB2004,FD2007,AG2010,TT2013,TT2013,TT201t,IC2015,YL2015}, and even a giant atom system which interacts with the waveguide via more than one connecting point~\cite{LG2017,LG2020,LD2021}. In these setups, the atomic population usually exhibits an oscillation behavior beyond the Markovian process.

In most of the previous studies about the retardation effect in waveguide system, the waveguide is usually theoretically considered as one dimensional. Therefore, the atom is resonantly coupled with only one flying photon mode in the waveguide. However, in the realistic physical system, the waveguide can never be one dimensional. Therefore, it is productive to investigate the effect of the finite cross section.

To tackle this issue, we here discuss the dynamics of two-atom system, which couples to a waveguide with rectangle cross section~\cite{JF,JL2021,Lj2020}. The finite cross section area of the waveguide generates two effects, the first one is that it supports more than one ${\rm TM}$ modes while the second one is that whether the atom is centered or off centered in the waveguide will lead to dramatically different dynamical behavior. For example, as both of the atoms are centered in the waveguide, the dynamics is similar to that in one dimensional waveguide, and we observe the non-Markovian retardation effect. Meanwhile, we recover the Markovian process by deviating one of the atom to be off-centered. We also find a dark state when the connection of the two atoms is properly perpendicular to the axis of the waveguide, in which both of the atoms will retain some excitation even after the evolution time is tend to be infinity.

The rest of the paper is organized as follows. In Sec.~\ref{model1}, we illustrate our model and give the general amplitudes equations. In Sec.~\ref{onemode}, we discuss the non-Markovian dynamics when the two atoms are both centered in the waveguide. In Sec.~\ref{twomode}, we consider the situation that one of the atom is off-centered. In Sec.~\ref{dark}, we reveal a dark state mechanism when the connection of the atoms is perpendicular to the axis of the waveguide. In Sec.~\ref{con}, we arrive at the conclusion.

\section{Model and amplitudes equations}
\label{model1}
\begin{figure}
\centering
\includegraphics[width=1\columnwidth]{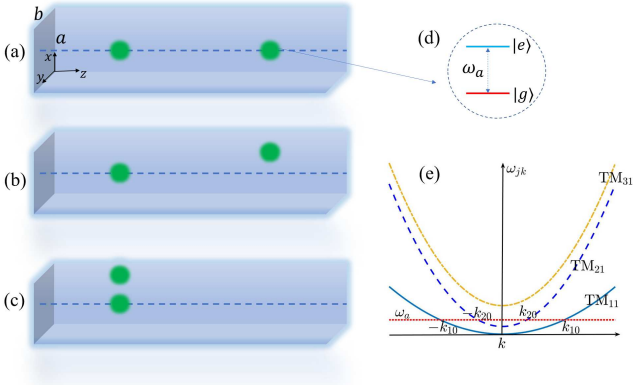}
\caption{Schematic illustration of two atoms couple to the waveguide with a $a\times b$ rectangle cross section. (a) The two atoms are both located in the middle axis of the waveguide. (b) One of the atom is off-centered. (c) The connection between the two atoms is perpendicular to the axis of the waveguide. (d) The energy-level diagram of the atoms. (e) The energy spectrum of the waveguide.}
\label{model}
\end{figure}

As schematically shown in Fig.~\ref{model} (a) (b) and (c), we consider a system composed by two two-level atoms, which couples to a common waveguide with a $a\times b$ rectangle cross section and being infinite in $z$ direction.  The two atoms are located at $\vec{r}_1=(x_1,y_1,z_1)$ and $\vec{r}_2=(x_2,y_2,z_2)$, respectively. The Hamiltonian of the coupled system is written as $H=H_0+H_I$ where ($\hbar=1$)
\begin{eqnarray}
H_0&=&\sum_{l=1}^{2}\omega_{a}\sigma^{+}_{l}\sigma^{-}_{l}
+\sum_j\int_{-\infty}^{\infty}dk\omega_{jk}a^{\dagger}_{jk}a_{jk},
\end{eqnarray}
describes the free energy of the atoms and the waveguide. Here, $\sigma^{+}_l=[\sigma^{-}_l]^{\dagger}=|e\rangle_l\langle g|$ is the Pauli operator of the $l$th atom. As shown in Fig.~\ref{model}(d), $\omega_a$ is the transition frequency between the atomic ground state $|g\rangle$ and excited state $|e\rangle$. $\omega_{jk}$ is the frequency of the travelling electromagnetic field mode in the waveguide. Here, the index $j$ denotes the electromagnetic field mode (see details below) and $k$ is the wave vector. $a_{jl}$ is the photon annihilation operator in the waveguide.

Within the rotating wave approximation, the interaction between the atoms and the waveguide is illustrated by the Hamiltonian
\begin{equation}
H_{I}=i\sum_{l=1}^{2}\sum_{j}\int_{-\infty}^{\infty}dk\frac{g_{jl}}
{\sqrt{\omega_{jk}}}\sigma^{-}_{l}a^{\dagger}_{jk}e^{ikz_{l}}+{\rm H.c.},
\end{equation}
where $z_l$ is the location of the $l$th atom in the $z$ direction. In this paper, we consider that the dipole moment of the atoms are along the $z$ direction, therefore, they are decoupled with the TE modes in the waveguide, that is, only the TM modes
are needed to be considered. For simplicity, we use a single notation $j$ to denote the TM modes. For simplicity, we denote $j=1$ for $m=1,n=1$, $j=2$ for $m=2,n=1$ and $j=3$ for $m=3,n=1$. Then, the atom-waveguide coupling strength and the dispersion relation of the waveguide are
\begin{eqnarray}
g_{jl}&=&\frac{\Omega_{j}\mu_j \sin(\frac{x_{l}m\pi}{a})\sin(\frac{y_{l}n\pi}{b})}{\sqrt{A\pi\epsilon_{0}}},\\
\omega_{jk}&=&\sqrt{\Omega_{j}^2+c^2k^2},\\\nonumber
\end{eqnarray}
respectively. Here $\Omega_{mn}=c\sqrt{(m\pi/a)^2+(n\pi/b)^2}$ is the cutoff frequency for a traveling wave for the ${\rm TM}_{mn}$ mode. $A=ab$ is the area of the rectangular cross section and $|\mu_{1}|=|\mu_{2}|=|\mu|$ is the magnitude of the transition dipole moment of the atom which is assumed be real. $c$ is the light velocity and $\epsilon_0$ is the permittivity of vacuum. The dispersion relation of the waveguide is demonstrate in Fig.~\ref{model} (e), and we set $\omega_a=(\Omega_1+\Omega_3)/2$, so that the atoms are large detuned from ${\rm TM}_{31}$ mode, but is resonant with ${\rm TM}_{11}$ and ${\rm TM}_{21}$ with certain wave vector.

Since the number of the quanta is conserved in our system, the wave function can be assumed as:
\begin{eqnarray}
|\psi(t)\rangle&=&e^{-i\omega_a t}[B_1(t)\sigma_1^{+}|G,0\rangle+B_2(t)\sigma_2^{+}|G,0\rangle]\nonumber\\
&&+\sum_{j}\int_{-\infty}^{\infty}dke^{-i\omega_{jk} t}B_{jk}(t)a^{\dagger}_{jk}|G,0\rangle,\\\nonumber
\end{eqnarray}
where $|G,0\rangle$ represents the state that both of the atoms are in their ground states while
the waveguide is in the vacuum state. $B_1(t)$ and $B_2(t)$ represent the excitation amplitudes for first and second atom while $B_{jk}$ is that for the $k$th mode of the waveguide for ${\rm TM}_{mn}$.  Based on the Sch\"{o}dinger equation, these amplitudes satisfy
\begin{eqnarray}
\dot B_1(t)&=&-\sum_{j}\int_{-\infty}^{\infty}dk\frac{g_{j1}B_{jk}(t)
e^{-i(\omega_{jk}-\omega_{a})t}e^{-ikz_{1}}}{\sqrt{\omega_{jk}}},\label{B1t}\\
\dot B_2(t)&=&-\sum_{j}\int_{-\infty}^{\infty}dk\frac{g_{j2}B_{jk}(t)
e^{-i(\omega_{jk}-\omega_{a})t}e^{-ikz_{2}}}{\sqrt{\omega_{jk}}},\label{B2t}\\
\dot B_{jk}(t)&=&\frac{(B_1(t)g_{j1}+B_2(t)g_{j2}e^{ikz_{0}})e^{ikz_{1}}
e^{i(\omega_{jk}-\omega_{a})t}}{\sqrt{\omega_{jk}}},\nonumber \\
\end{eqnarray}
where $z_0=z_2-z_1$. In the initial vacuum waveguide condition $B_{jk}(0)=0$, the excited amplitudes of the
waveguide can be obtained formally as
\begin{equation}
B_{jk}(t)=\int_{0}^{t}d\tau\frac{e^{ikz_{1}}}{\sqrt{w_{jk}}}
[g_{j1}B_{1}(\tau)+g_{j2}B_{2}(\tau)e^{ikz_{0}}]e^{i(w_{jk}-w_{a})\tau}.
\end{equation}
Substituting $B_{jk}(t)$ into Eqs.~(\ref{B1t}) and (\ref{B2t}), the retardation differential equations for the atomic amplitudes are obtained as
\begin{eqnarray}
&&(\partial_{t}+\sum_{j}\frac{g_{j1}^2\pi}{\omega_{a}v_{j}})B_1(t)=\nonumber\\
&&-\sum_{j}\frac{g_{j1}g_{j2}}{\omega_{a}v_{j}}B_2(t-\frac{d}{v_j})e^{ik_{j0}d}
\Theta(t-\frac{d}{v_j}),\label{A1}\\
&&(\partial_{t}+\sum_{j}\frac{g_{j2}^2\pi}{\omega_{a}v_{j}})B_2(t)\nonumber
=\\
&&-\sum_{j}\frac{g_{j1}g_{j2}}{\omega_{a}v_{j}}B_1(t-\frac{d}{v_j})e^{ik_{j0}d}
\Theta(t-\frac{d}{v_j})\label{A2}.
\end{eqnarray}
In the above equations, $k_{j0}=\sqrt{\omega_{a}^{2}-\Omega_{j}^{2}}/c$ is the wave vector of the waveguide mode which is resonant with the atoms and $d=|z_{0}|$ is the distance of the two atoms in the $z$ direction. The corresponding group velocity $v_j$ is
\begin{equation}
v_j=\frac{d\omega_{jk}}{dk}|_{k=k_{j0}}=\frac{c\sqrt{\omega_{a}^2-\Omega_{j}^2}}{\omega_{a}}.
\end{equation}

We emphasize that Heaviside unit step function $\Theta(x)$, which is defined as $\Theta(x)=1$ for $x>0$ and $\Theta(x)=0$ for $x\leq 0$, represents the non-Markovian retardation effects, since $d/v_j$ corresponds to the time needed for the photon propagating for one atom to the other.

\section{Two atoms centered in the waveguide}
\label{onemode}
As shown in Fig.~1(a), we now consider that the two atoms are both located in the middle of the waveguide, that is, $\vec{r}_1=(a/2,b/2,z_1)$ and $\vec{r}_2=(a/2,b/2,z_2)$. A direct observation shows that $g_{2i}=0$ for $i=1,2$. Therefore, the atoms are coupled to the ${\rm TM_{11}}$ mode in the waveguide and the coupling strengths are obtained as
\begin{equation}
g_{11}=g_{12}=\frac{\Omega_{1}\mu}{\sqrt{A\pi\epsilon_{0}}}.
\label{strength}
\end{equation}
As a result, the amplitude equation in Eqs.~(\ref{A1}) and (\ref{A2}) becomes
\begin{subequations}
\begin{eqnarray}
(\partial_{t}+\gamma_{11})B_1(t)=-\gamma_{11}B_2(t-\tau_{1})e^{ik_{10}d}\Theta(t-\tau_{1}),\nonumber \\ \\
(\partial_{t}+\gamma_{11})B_2(t)=-\gamma_{11}B_1(t-\tau_{1})e^{ik_{10}d}\Theta(t-\tau_{1}),\nonumber \\
\end{eqnarray}
\label{RE}
\end{subequations}
where $\gamma_{11}=g_{11}'^{2}\pi/v_{1}$ is the effective decay rate of the atoms (equal for each atom), $g_{11}'=g_{11}/\sqrt{\omega_{a}}$ is the renormalized coupling strength under the Weisskopf-Wigner approximation~\cite{MO1997}. $\tau_1=d/v_1$ is the delay time for the photon with group velocity $v_1$ travelling from one atom to the other in the waveguide.
\begin{figure}
\centering
\includegraphics[width=1\columnwidth]{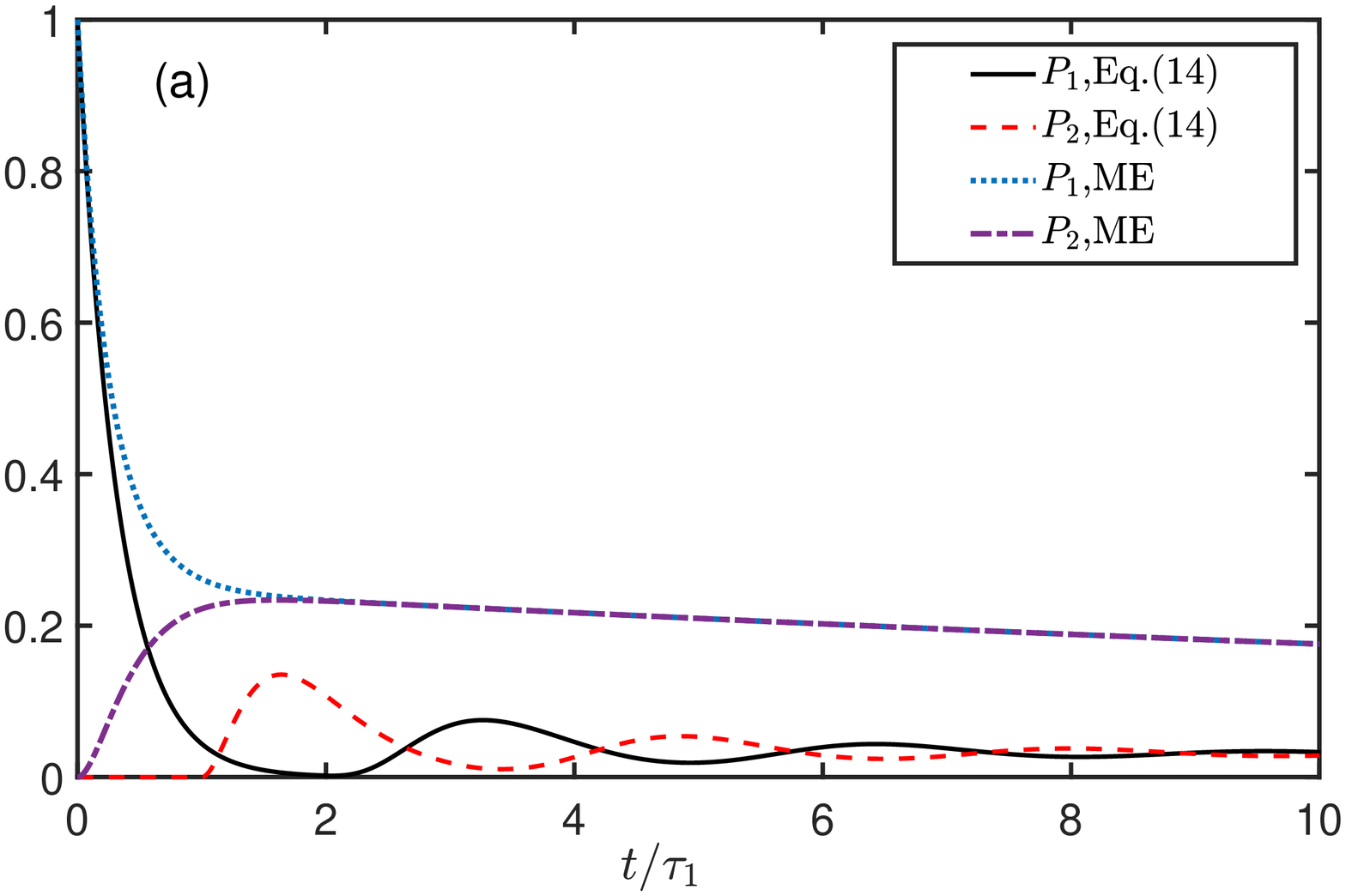}
\includegraphics[width=1\columnwidth]{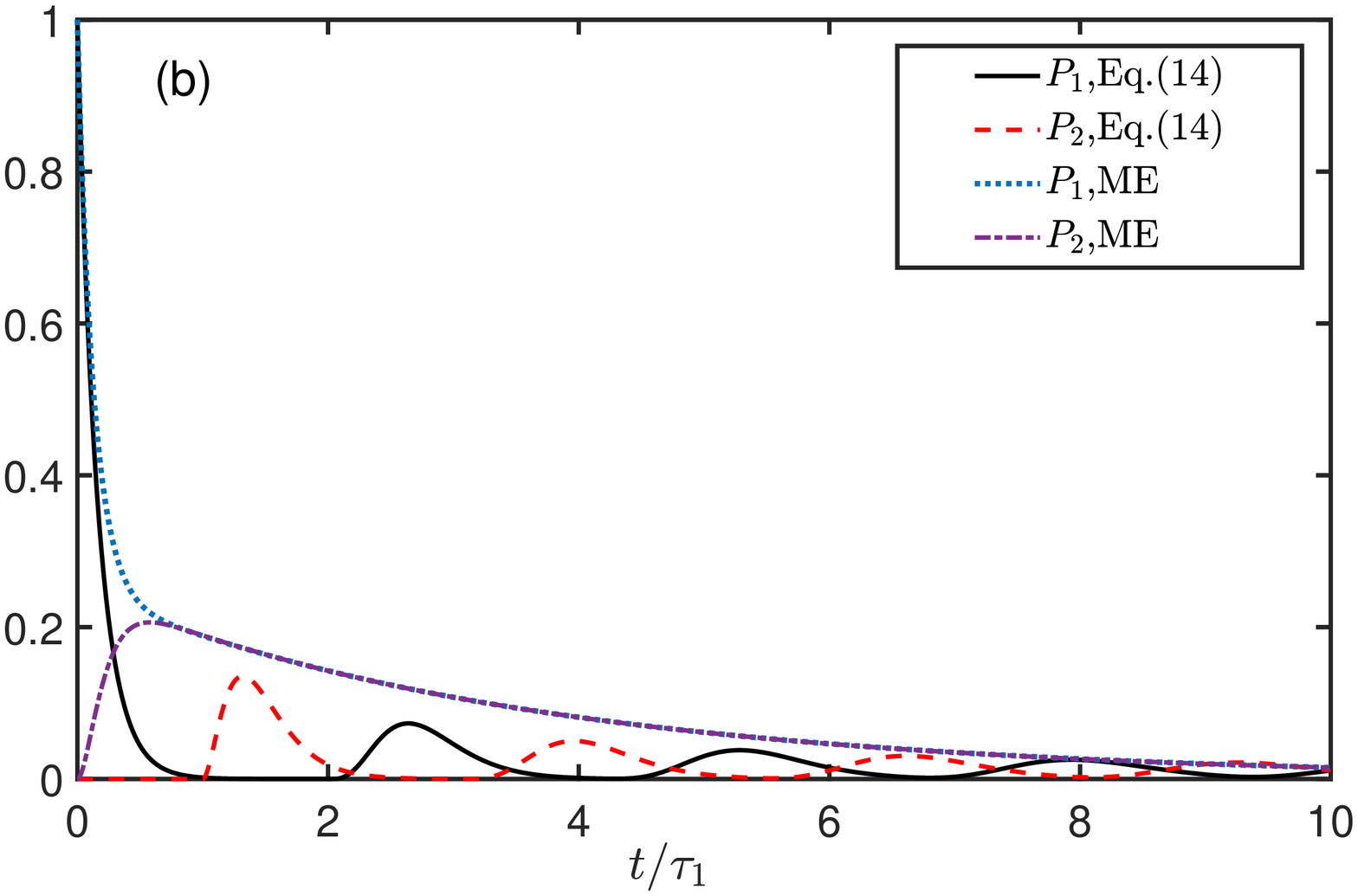}
\caption{Dynamical evolution of the atomic populations based on the retardation differential equations in Eq.~(\ref{RE}) and ME in Eq.~(\ref{master}). The parameters are set as
(a) $d=12/k_{j0}$, (b) $d=24/k_{j0}$ and $\Omega_1a/(c\pi)=\sqrt{5},b=a/2,\omega_a=(\Omega_1+\Omega_3)/2,g_{11}'=0.08\Omega_1$.}
\label{onem}
\end{figure}
In the viewpoint of quantum open system, the electromagnetic field in the waveguide serves as an environment, which induces the dissipation and indirect interaction between the atoms. Under the Born-Markovian approximation, the dynamics of the two atoms are governed by the master equation (ME)
\begin{equation}
\dot\rho=-i[\mathcal{H}_1,\rho]+\sum_{i,j=1}^{2}\frac{\Gamma_{ij}}{2}
(2\sigma_{j}^{-}\rho\sigma_{i}^{+}-\sigma_{i}^{+}\sigma_{j}^{-}\rho
-\rho\sigma_{i}^{+}\sigma_{j}^{-}),
\label{master}
\end{equation}
where the Hamilton $\mathcal{H}_1$ between the two atoms reads
\begin{equation}
\mathcal{H}_1=\sum_{i=1}^{2}\omega_{a}(\sigma_i^{+}\sigma_i^{-})
+\sum_{i,j=1}^{2}\frac{U_{ij}}{2}(\sigma_{i}^{+}\sigma_{j}^{-}+\sigma_{i}^{-}\sigma_{j}^{+}).
\end{equation}
where $\Gamma_{ij}=2{\rm Re}(A_{ij})$ is the two-atom collective decay rate and $U_{ij}=2{\rm Im}(A_{ij})$ is the waveguide induced interaction between the atoms. Here, we have set
\begin{equation}
A_{ij}=\frac{\pi g^{2}e^{ik|z_i-z_j|}}{v_1}.
\end{equation}

The dynamics of the system, which is characterized by the atomic population $P_i=\langle\sigma_i^{+}\sigma_i^{-}\rangle=|B_i|^2$ for $i=1,2$ is shown in Fig.~\ref{onem} for different atomic distance. Here, the system is initially prepared in the product state $|\psi(0)\rangle=\sigma_1^{+}|G,0\rangle$, in which the first atom is in the excited state, the second atom is in the ground state while the waveguide is in the vacuum state.

In Fig.~\ref{onem}(a), we consider the situation with $d=12/k_{j0}$, in which the ME yields a monotonous decay for $P_1$ and an increase-decrease transition for $P_2$. However, the results based on the Eqs.~(\ref{RE}) reveals the Non-Markovian nature of the system which is induced by the retardation effect during the photon propagation in the waveguide. For example, at the moment $t=\tau_1$, the emitted photon by the first atom arrives at the second atom and excites it, so $P_2$ acquires a non-zero value. Then, it also emits photon, which in turn arrives at the first atom during another time interval $\tau_1$, and the decreasing population $P_1$ revivals along with the reabsorption of photon. Repeating such photon emitting, propagation and absorbing process, both of the population $P_1$ and $P_2$ oscillates with period $\tau_1$. Due to the waveguide induced dissipation for the two-atom system, the populations will achieve zero after a sufficient long time.  The similar behavior can also be found for a large atomic distance $d=24/k_{j0}$ as shown in Fig.~\ref{onem}(b). Comparing with the former situation, we find that the population will undergo a longer time to stay at the zero value (see the black solid curve nearby $t/\tau_1=2$ and the red dashed curve nearby $t/\tau_1=3$ ) due to the longer
retardate time.

\section{Effects of $ \text{TM}_{21}$ mode}
\label{twomode}

Now, we consider that the second atom is off-centred from the waveguide as shown in Fig.~\ref{model}(b), that is $\vec{r}_1=(a/2,b/2,z_1)$ and $\vec{r}_2=(a/2+\Delta x,b/2,z_2)\,(0<\Delta x<a/2)$. An immediate result is the change of atom-waveguide coupling strength, which yields
\begin{equation}
g_{11}=\frac{\Omega_{1}\mu}{\sqrt{A\pi\epsilon_{0}}},g_{12}=g_{11}\cos(\frac{\Delta x\pi}{a}),g_{12}'=\frac{g_{12}}{\sqrt{\omega_{a}}},
\end{equation}
More interesting, it is non-trivial that the second atom couples to the ${\rm TM}_{21}$ mode simultaneously besides ${\rm TM}_{11}$ mode and the coupling strength reads
\begin{equation}
g_{22}=\frac{\Omega_{2}\mu}{\sqrt{A\pi\epsilon_{0}}}\sin(\frac{2\Delta x\pi}{a}).
\end{equation}
\begin{figure}
\centering
\includegraphics[width=0.95\columnwidth]{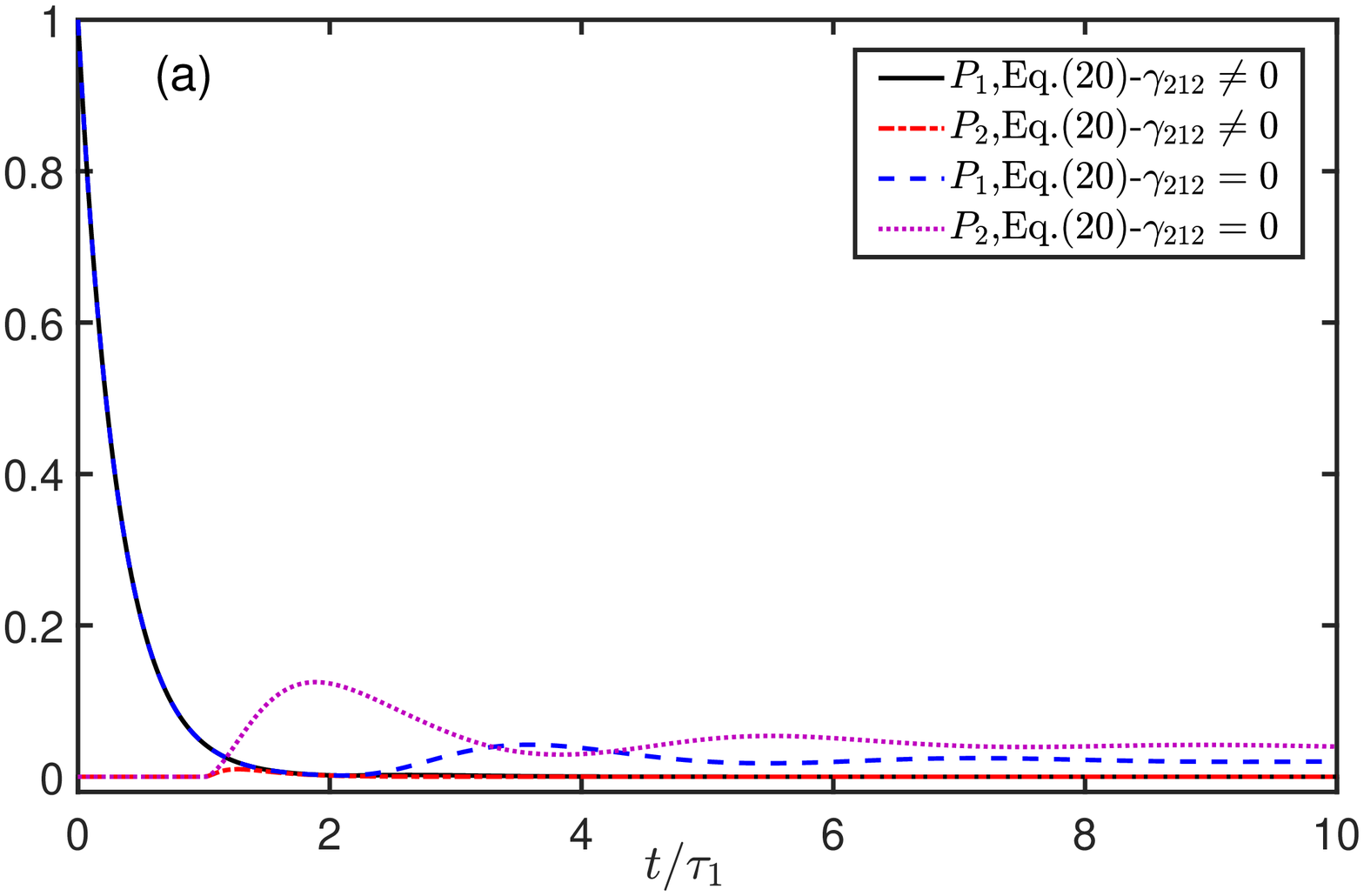}
\includegraphics[width=0.95\columnwidth]{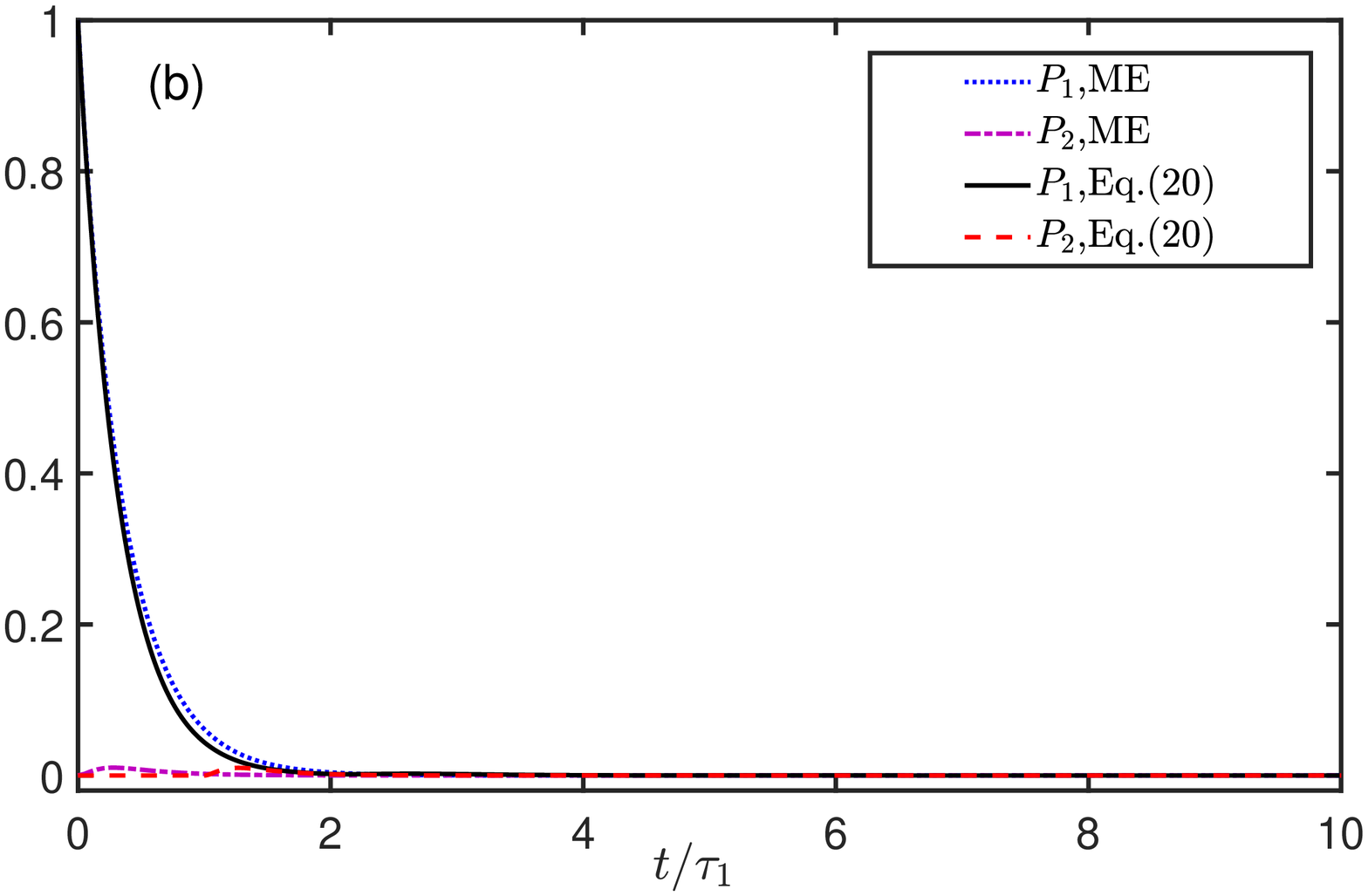}
\caption{Dynamical evolution of the atomic populations when the second atom is off centred from the middle axis of the waveguide. (a) The comparison when the ${\rm TM}_{21}$ mode is considered ($\gamma_{212}\neq0$) or neglected ($\gamma_{212}=0$).
(b) The comparison of the results based on the retardation differential equations in Eq.~(\ref{newmode}) and ME in Eq.~(\ref{newmodemaster}). The parameters are set as
$d=12/k_{j0}$,  $\Delta x=a/4$, $\Omega_1a/(c\pi)=\sqrt{5},b=a/2,\omega_a=(\Omega_1+\Omega_3)/2,g_{11}'=0.08\Omega_1$.}
\label{twomai}
\end{figure}
As a result, the retardation differential equations for the atomic amplitudes becomes
\begin{subequations}
\begin{eqnarray}
(\partial_{t}+\gamma_{11})B_1(t)&=&-\gamma_{12}B_2(t-\tau_{1})e^{ik_{10}d}
\Theta(t-\tau_{1}),\nonumber\\
\\
(\partial_{t}+\gamma_{22}+\gamma_{212})B_2(t)&=&-\gamma_{12}B_1(t-\tau_{1})
e^{ik_{10}d}\Theta(t-\tau_{1})\nonumber,\\
\end{eqnarray}
\label{newmode}
\end{subequations}
where
\begin{eqnarray}
\gamma_{11}&=&\frac{g_{11}'^{2}\pi}{v_{1}},\,
\gamma_{12}=\frac{g_{11}'g_{12}'\pi}{v_{1}},\label{gamma1}\\
\gamma_{22}&=&\frac{g_{12}'g_{12}'\pi}{v_{1}},\,
\gamma_{212}=\frac{g_{22}'^2\pi}{v_2}.\label{gamma2}
\end{eqnarray}
with $g_{mn}'=g_{mn}/\sqrt{\omega_a}$. It is clear that $\gamma_{11}, \gamma_{12}$ and $\gamma_{22}$ come from the coupling to the ${\rm TM}_{11}$ mode while $\gamma_{212}$ comes from the effect of ${\rm TM}_{21}$ mode.

To demonstrate the effect of coupling to ${\rm TM}_{21}$ mode of the second atom, we plot the atomic populations for $\gamma_{212}=0$ and $\gamma_{212}\neq0$ based on Eq.~(\ref{newmode}) in Fig.~3(a) with the initial state being same with that in the last section.
When $\gamma_{212}$ is considered to be zero, that is, the ${\rm TM}_{21}$ mode is neglected, both $P_1$ and $P_2$ will experience the oscillation, which is similar to the situation when the two atoms are both centered in the waveguide, and the difference comes from the modification of coupling strength between the second atom and the waveguide due to the derivation. However, as the effect of the ${\rm TM}_{21}$ is taken into consideration ($\gamma_{212}\neq0$), $P_1$ experiences an exponential decay and $P_2$  nearly stays in the ground state all the time. Therefore, the ${\rm TM}_{21}$ mode provides a new dissipation channel for the second atom, which prevents its excitation.

Similar to the discussion in the last section, we can also obtain the ME under the Markovian approximation, which yields
\begin{eqnarray}
\dot\rho&=&-i[\mathcal{H}_2,\rho]+\sum_{i,j=1}^{2}\frac{\Gamma_{ij}'}{2}
(2\sigma_{j}^{-}\rho\sigma_{j}^{+}-\sigma_{i}^{+}\sigma_{j}^{-}\rho
-\rho\sigma_{i}^{+}\sigma_{j}^{-})\nonumber\\
&&+\gamma_{212}(2\sigma_{2}^{-}\rho\sigma_{2}^{+}
-\sigma_{2}^{+}\sigma_{2}^{-}\rho-\rho\sigma_{2}^{+}\sigma_{2}^{-}).
\label{newmodemaster}
\end{eqnarray}
Here, the last term represents the dissipation of the second atom induced by the ${\rm TM}_{21}$ mode in the waveguide. The Hamilton $\mathcal{H}_2$ for the interaction between the two atoms reads
\begin{equation}
\mathcal{H}_2=\sum_{l=1}^{2}\omega_{a}(\sigma_l^{+}\sigma_l^{-})
+\sum_{i,j=1}^{2}\frac{U_{ij}'}{2}(\sigma_{i}^{+}\sigma_{j}^{-}+\sigma_{i}^{-}\sigma_{j}^{+}),
\end{equation}
where
\begin{equation}
U_{ij}'=\gamma_{12}\sin|z_i-z_j|,\Gamma_{ij}'=\frac{2\pi g_{1i}g_{1j}\cos|z_{i}-z_{j}|}{c\sqrt{\omega_{a}^{2}-\Omega_{1}^{2}}}.
\end{equation}

In Fig.~\ref{twomai}(b), we show the agreement of the results between the retardation differential equations and ME. When the ${\rm TM}_{21}$ mode in considered, the second atom immediately decays after it is excited by the photon emitted by the first atom, so that we can barely observe the oscillation. Meanwhile, the photon emitted by the first atom propagates via ${\rm TM}_{11}$ mode can not be reflected by the second atom due to its dissipation via ${\rm TM}_{21}$ mode and therefore the ME works well and $P_1$ exhibits an exponential decay.

\section{Dark state}
\label{dark}

In the above sections, we have considered the situation with $d=|z_1-z_2|\neq0$, in which the dynamics of the system is demonstrated by the retardation differential equations.  Another interesting situation is that the connection between the two atoms is perpendicular to the axis of the waveguide. We first consider the case illustrated in Fig.~\ref{model} (c), where both of the atoms are located in the position $z=z_0$ but the second atom is deviated from the first one in the $x$ direction, that is,  $\vec{r}_1=(a/2,b/2,z_0)$ and $\vec{r}_2=(a/2+\Delta x,b/2,z_0)\,(0<\Delta x<a/2)$. As a result, there is no retardation effect and the amplitudes satisfy the differential equations
\begin{subequations}
\begin{eqnarray}
(\partial_{t}+\gamma_{11})B_1(t)&=&-\gamma_{12}B_2(t),\nonumber \\
\\
(\partial_{t}+\gamma_{22}+\gamma_{212})B_2(t)&=&-\gamma_{12}B_1(t),\nonumber\\
\end{eqnarray}
\label{per1}
\end{subequations}
where the parameters $\gamma_{11},\gamma_{22}, \gamma_{212}$ and $\gamma_{12}$ are same with those given in Eqs.~(\ref{gamma1}) and (\ref{gamma2}). Correspondingly, the Markovian master equation becomes
\begin{eqnarray}
\dot\rho&=&-i[\mathcal{H}_3,\rho]+\sum_{i,j=1}^{2}\frac{\Gamma_{ij}^{*}}{2}
(2\sigma_{j}^{-}\rho\sigma_{i}^{+}-\sigma_{i}^{+}\sigma_{j}^{-}\rho
-\rho\sigma_{i}^{+}\sigma_{j}^{-})\nonumber\\
&&+\gamma_{212}(2\sigma_{2}^{-}\rho\sigma_{2}^{+}
-\sigma_{2}^{+}\sigma_{2}^{-}\rho-\rho\sigma_{2}^{+}\sigma_{2}^{-}),
\label{mep}
\end{eqnarray}
where the Hamiltonian
\begin{equation}
\mathcal{H}_3=\sum_{l=1}^{2}\omega_{a}(\sigma_l^{+}\sigma_l^{-}),
\end{equation}
implies that the two atoms do not coherently couple to each other. However, the nonzero value of
\begin{equation}
\Gamma_{ij}^{*}=\frac{2\pi g_{1i}g_{1j}}{c\sqrt{\omega_{a}^{2}-\Omega_{1}^{2}}}.
\end{equation}
indicates that they will undergo a collective dissipation to the waveguide.
\begin{figure}
\centering
\includegraphics[width=0.95\columnwidth]{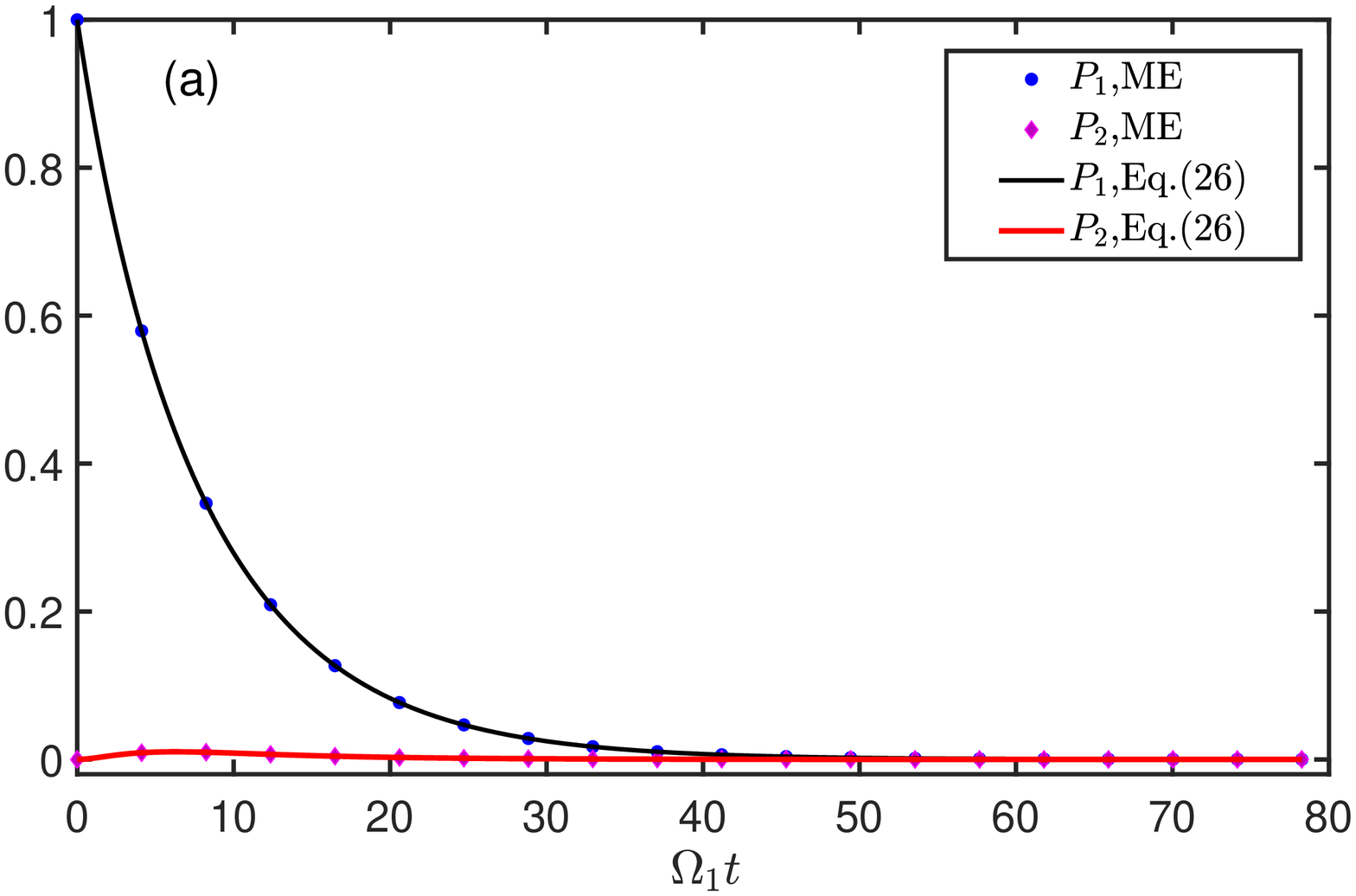}
\includegraphics[width=0.95\columnwidth]{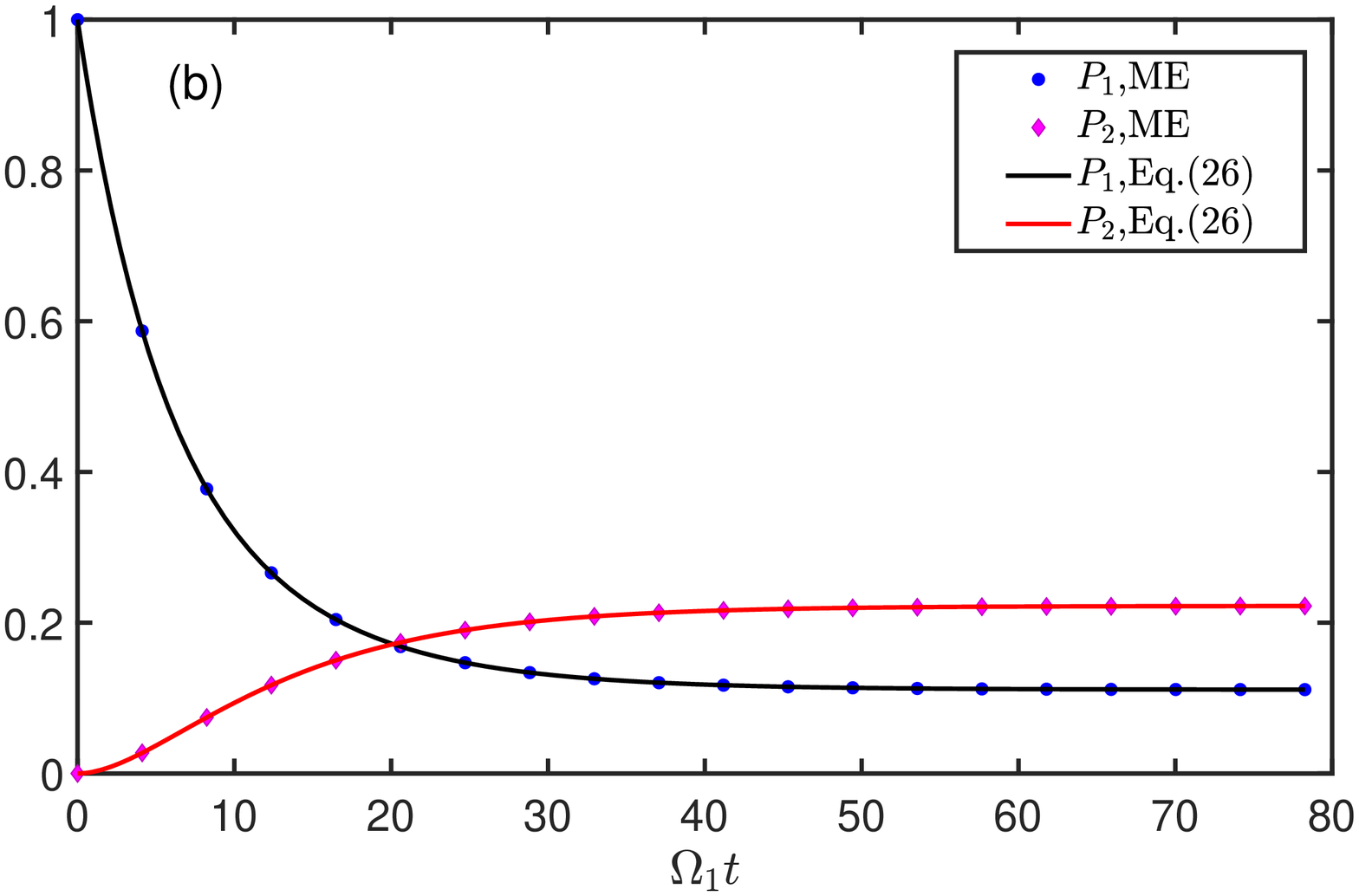}
\caption{Dynamical evolution of the atomic populations when the connection of the two atoms are perpendicular to the axis of the waveguide. The parameters are set as $d=12/k_{j0}$, $\Omega_1a/(c\pi)=\sqrt{5},b=a/2,\omega_a=(\Omega_1+\Omega_3)/2,g_{11}'=0.08\Omega_1$. (a) $\Delta x=a/4$ and (b) $\Delta y=b/4$.}
\label{twomai}
\end{figure}
Fig.~4(a) shows the comparison between Eqs.~(\ref{per1}) and ME for dynamical evolution of the atomic populations. The absence of the retardation yields agreement of the two results as shown in the figure. Furthermore, $P_1$ experiences an exponential decay and  $P_2$ nearly keeps zero during the time evolution, therefore, the second atom is nearly frozen in the ground state.

The above dynamical process can be broken if the second atom is deviated from the first one in the $y$ direction instead of $x$ direction, that is, $\vec{r}_1=(a/2,b/2,z_0)$ and $\vec{r}_2=(a/2,b/2+\Delta y,z_0)\,(0<\Delta y<b/2)$. In this case, the amplitudes equations and the ME are same with Eqs.~(\ref{per1}) and (\ref{mep}), respectively, the only difference is $\gamma_{212}=0$ since the second atom is decoupled from the ${\rm TM}_{21}$ mode.  As shown in Fig.~4(b), the ME describes the dynamics of the system perfectly  and the atomic populations will achieve an nonzero fixed value as the evolution time $t$ tends to be infinite. It means that the system finally reaches a dark state which protects the atoms from decaying to the ground state. The underlying physics can be extracted from the effective interaction Hamiltonian, which is simplified as
\begin{equation}
H_{I}=i\int_{-\infty}^{\infty}dk\frac{1}{\sqrt{\omega_{1k}}}\left[(g_{11}
\sigma^{-}_{1}+g_{12}\sigma^{-}_{2})a^{\dagger}_{1k}e^{ikz_0}-{\rm H.c.}\right].
\end{equation}
As a result, the dark state $|D\rangle$ which satisfies $H_{I}|D\rangle=0$ can be expressed as
\begin{equation}
|D\rangle=\frac{1}{\sqrt{g_{12}^2+g_{11}^2}}({g_{12}\sigma_1^{+}-g_{11}\sigma_2^{+}})|G,0\rangle.
\end{equation}
Therefore, we have
\begin{equation}
\label{dar}
\frac{P_{1}(t=\infty)}{P_{2}(t=\infty)}=\frac{g_{12}^{2}}{g_{11}^{2}}=\cos^{2}(\frac{y\pi}{b}),
\end{equation}
which coincides with results given in Fig.~4(b).

\section {conclusion}
\label{con}
In this paper, we investigate the time evolution of a two-atom system which couples to a waveguide with rectangle cross section. We find that the dynamics of the system can be controlled by adjusting the
relative location of the two atoms in a on-demand manner. Similar to the modeled waveguide system, where the effect of cross section is neglected, the dynamics exhibits an obvious non-Markovian retardation character when the atoms are both located on the middle axis of the waveguide, since they interact with the same mode in the waveguide. This mode not only induces the dissipation but also plays a data bus to indirectly couple the two atoms. As one of the atom is off-centered, an additional mode in the waveguide acts as pure dissipation environment, which erodes the retardation effect and therefore the Markovian master equation will capture the main physics in an analytical way. More interestingly, when the connection of the atoms are perpendicular to the axis of the waveguide, we find a dark state mechanism which prevents the complete decay of the system and is therefore of potential application in quantum information processing.

\begin{acknowledgments}
We thank Dr. L. Du for warm help. This work is supported by National Key R$\&$D Program of China (No. 2021YFE0193500), and the National Natural Science Foundation of China (No. 11875011).
\end{acknowledgments}


\begin{thebibliography}{99}

\bibitem{DR2017}D. Roy, C. M. Wilson and O. Firstenberg,
Rev. Mod. Phys. {\bf 89}, 021001 (2017).

\bibitem{XG2017}X. Gu, A. F. Kockum, A. Miranowicz, Y.-x. Liu and F. Nori,
Phys. Rep. {\bf 718}, 1 (2017).


\bibitem{JT2005}J. T. Shen and S. Fan, Phys. Rev. Lett. {\bf 95}, 213001 (2005).

\bibitem{DE2007}D. E. Chang, A. S. S{\o}rensen, E. A. Demler and M. D. Lukin,
Nat. Phys. {\bf 3}, 807 (2007).

\bibitem{LZ2008}L. Zhou, Z. R. Gong, Y. X. Liu, C. P. Sun and F. Nori, Phys.
Rev. Lett. {\bf 101}, 100501 (2008).


\bibitem{AA2010} A. A. Abdumalikov Jr., O. Astafiev, A. M. Zagoskin, Y. A.
Pashkin, Y. Nakamura and J. S. Tsai,  Phys. Rev. Lett. {\bf 104}, 193601 (2010).

\bibitem{IC2011}I.-C. Hoi, C. M. Wilson, G. Johansson, T. Palomaki, B.
Peropadre and P. Delsing, Phys. Rev. Lett. {\bf 107}, 073601
(2011).

\bibitem{LZ2013}L. Zhou, L. P. Yang, Y. Li and C. P. Sun, Phys. Rev. Lett. {\bf 111},
103604 (2013).

\bibitem{Wang2014}Z. H. Wang, L. Zhou, Y. Li and C. P. Sun, Phys. Rev. A {\bf 89}, 053813 (2014).

\bibitem{IS2014}I. Shomroni, S. Rosenblum, Y. Lovsky, O. Bechler, G.
Guendelman and B. Dayan, Science {\bf 345}, 903 (2014).

\bibitem{CH2018}C.-H. Yan, Y. Li, H. Yuan and L. F. Wei, Phys. Rev. A {\bf 97},
023821 (2018).

\bibitem{YL2022}Y.-l. Ren, S.-l. Ma, J.-k. Xie, X.-k. Li, M.-t. Cao, and F.-l. Li, Phys. Rev. A {\bf 105}, 013711 (2022).

\bibitem{KS2012} K. Stannigel, P. Rabl and P. Zoller, New J. Phys. {\bf 14}, 063014
(2012).

\bibitem{AG2013} A. Gonz\'{a}lez-Tudela and D. Porras, Phys. Rev. Lett. {\bf 110}, 080502
(2013).

\bibitem{CG2013} C. Gonzales-Ballestero, F. J. Garcia-Vidal and E.Moreno, New
J. Phys. {\bf 15}, 073015 (2013).

\bibitem{HP2015}H. Pichler, T. Ramos, A. J. Daley and P. Zoller, Phys. Rev. A
{\bf 91}, 042116 (2015).

\bibitem{GC2016}G. Calaj\'{o}, F. Ciccarello, D. Chang and P. Rabl, Phys. Rev. A
{\bf 93}, 033833 (2016).

\bibitem{AG2017}A. Gonz\'{a}lez-Tudela and J. I. Cirac
Phys. Rev. A {\bf 96}, 043811 (2017).

\bibitem{FG2018}F. Galve and R. Zambrini,
Phys. Rev. A {\bf 97},  033846 (2018).

\bibitem{HZ2019}H. Z. Shen, S. Xu, H. T. Cui and X. X. Yi,
Phys. Rev. A {\bf 99}, 032101 (2019).

\bibitem{EK2021}E. Kim, X. Zhang, V. S. Ferreira, J.Banker, J. K. Iverson, A. Sipahigil, M. Bello, A. G.-Tudela, M. Mirhosseini and O. Painter, Phys. Rev. X {\bf 11},  011015 (2021).


\bibitem{FT1970}F. T. Arecchi and E. Courtens, Phys. Rev. A {\bf 2}, 1730
(1970).

\bibitem{PW1974}P. W. Milonni and P. L. Knight, Phys. Rev. A {\bf 10}, 1096
(1974).

\bibitem{Qu2012}Q.-ul-Ain Gulfam, Z. Ficek and J. Evers, Phys. Rev. A {\bf 86}, 022325 (2012).

\bibitem{HP2016}H. Pichler and Peter Zoller, Phys. Rev. Lett. {\bf 116}, 093601(2016).

\bibitem{KS2020} K. Sinha, P. Meystre, E. A. Goldschmidt, F. K. Fatemi, S. L.
Rolston and P. Solano, Phys. Rev. Lett. {\bf 124}, 043603 (2020).


\bibitem{JE2001} J. Eschner, C. Raab, F. Schmidt-Kaler and R. Blatt, Nature {\bf 413}, 495 (2001).

\bibitem{UD2002} U. Dorner and P. Zoller, Phys. Rev. A {\bf 66}, 023816 (2002).

\bibitem{PB2004} P. Bushev, A. Wilson, J. Eschner, C. Raab, F. Schmidt-Kaler, C. Becher and R. Blattet, Phys. Rev. Lett. {\bf 92}, 223602 (2004).

\bibitem{FD2007} F. Dubin,  M. Mukherjee, C. Russo, J. Eschner and R. Blatt, Phys. Rev. Lett. {\bf 98}, 183003 (2007).

 \bibitem{AG2010}A.  Glaetzle,  K.  Hammerer,  A.  Daley,  R.  Blatt and  P.  Zoller,
 Optics Communications {\bf 283}, 758 (2010)

\bibitem{TT2013} T. Tufarelli, F. Ciccarello and M. S. Kim, Phys. Rev.
A {\bf 87}, 013820 (2013).

\bibitem{TT201t}T. Tufarelli, M. S. Kim and F. Ciccarello, Phys. Rev. A {\bf 90}, 012113
(2014).

\bibitem{IC2015}I.-C. Hoi, A. F. Kockum, L. Tornberg, A. Pourkabirian, G. Johansson, P. Delsing and C. M. Wilson, Nat. Phys.  {\bf 11}, 1045 (2015).

\bibitem{YL2015}Y.-L. L. Fang and Harold U. Baranger
Phys. Rev. A {\bf 91}, 053845 (2015).

\bibitem{LG2017}L. Guo, A. L. Grimsmo, A. F. Kockum, M. Pletyukhov and G.
Johansson,  Phys. Rev. A {\bf 95}, 053821 (2017).

\bibitem{LG2020}L. Guo, A. F. Kockum, F. Marquardt and G. Johansson, Phys. Rev. Research {\bf 2}, 043014 (2020).

\bibitem{LD2021}L. Du, M.-R. Cai, J.-H. Wu, Z. Wang and Y. Li, Phys. Rev. A {\bf 103}, 053701 (2021).


\bibitem{JF}J.-F. Huang, T. Shi, C. P. Sun and F. Nori,
Phys. Rev. A {\bf 88}, 013836 (2013).

\bibitem{JL2021}J. Li, L. Hu, J. Lu and L. Zhou, Chin. Phys. B {\bf 30}, 090307 (2021).

\bibitem{Lj2020}L. Hu, G. Lu, J. Lu and L. Zhou, Quantum Inf. Process {\bf 19}, 81 (2020).

\bibitem{MO1997}M. O. Scully and M. S. Zubairy, \textit{Quantum Optics}, 1st ed. (Cambridge University Press, 1997), Page 206.

\end{thebibliography}
\end{document}